\documentclass{article}
\usepackage{spconf,amsmath,graphicx,hyperref}
\usepackage{booktabs}
\usepackage{multirow}
\title{Mic Drop: Bridging Device Gaps in Acoustic Scene Classification via Unsupervised Domain Adaptation}
\name{Abhishek Dileep, Naman Pratap Singh, Himanshu Yadav, Padmanabhan Rajan}
\address{School of Computing and Electrical Engineering, IIT Mandi \\
Mandi, Himachal Pradesh, India \\
\{\texttt{s23103, s25088, d25057\}@students.iitmandi.ac.in} \\
\texttt{padman@iitmandi.ac.in}
}
\begin{document}
\ninept
\maketitle

\begin{abstract}
Domain adaptation is a well-studied problem in machine learning. This topic is less studied for audio data. In this paper, we evaluate well-known domain adaptation methods from computer vision, including domain adapted neural network, maximum classifier discrepancy and transferable vision transformers for audio spectrograms. Our experiments show that many of the methods carry forward well to spectrograms, though some do not. We also propose a knowledge distillation based method for multi-target domain adaptation. Multiple domain adapted teachers distill information into a lighter student model, with the result that the student retains most of the original domain information, and generalizes to the new domains. Our experiments on acoustic scene classification data using devices as domains highlight our observations.

\end{abstract}
\begin{keywords}
Acoustic scene classification, Domain adaptation, Knowledge distillation, Adversarial training
\end{keywords}
\section{Introduction}
\label{sec:intro}

Machine learning (ML) based classifiers frequently perform poorly under domain mismatch ~\cite{ASC_with_ADA,ASC_with_WGAN}. Domain adaptation of ML models is widely studied, and several methods have been proposed in the literature for handling domain shift. Most of the studies are centered around image-based domain adaptation algorithms, on standard domain adaptation datasets such as office-home, Visda ~\cite{visda,officehome}.  Not many studies have attempted an explicit domain adaptation for audio signals. In this paper, we experiment with a variety of domain adaptation methods, known to work for images, and
evaluate them systematically for audio classification tasks. In addition, we propose a knowledge-distillation based method where domain knowledge is transferred from multiple domain-specific teachers to a single domain-invariant student  for multi-target domain adaptation ~\cite{MTDA}.

We perform the study under Unsupervised Domain Adaptation (UDA), where we have labeled data from a source domain and unlabeled data from a target domain. Among the several well-known solutions to UDA, we investigate Gradient Reversal Layer (GRL), also called Domain-Adversarial Neural Networks (DANN) ~\cite{DANN}, Maximum Classifier Discrepancy (MCD) ~\cite{MCD} and Transferable Vision Transformers (TVT) ~\cite{TVT}. We evaluate domain adaptation for acoustic scene classification (ASC), where different domains are represented by different recording devices. We investigate two pipelines, one based on audio spectrogram transformer (AST) ~\cite{AST}, and the other on the MobileNet convolutional neural network (CNN) ~\cite{mobilenetv3}.

%In many situations, there is a need to adapt to multiple target domains.
Multi-target domain adaptation (MTDA) is achieved by using techniques including discrepancy minimization and feature disentanglement \cite{gholami2020unsupervised}, and multi-adversarial domain alignment \cite{pei2018multi}.
Another commonly used method for  MTDA is via knowledge distillation (KD.) Existing work distills CNN teacher models into student models \cite{nguyen2021unsupervised}, \cite{takeyama2021robust}.
%distills multiple CNN teachers into one CNN student. %This distillation consolidates multi-target knowledge into a single deployable model. Similar KD-based MTDA remains underexplored for acoustic scene classification.
%Existing ASC-specific KD work also distills a CNN teacher into a CNN student \cite{takeyama2021robust}.
In this work, we leverage a pool of single-target AST-based domain-adapted models (called teacher models) and transfer the information to a MobileNet-based  student model. The aim is to make the student model obtain information from all teacher models, and thereby achieve multi-target domain adaptation.

The major contributions of this paper are follows:
\begin{enumerate}
\item We provide a comprehensive study of domain adaptation techniques applied to device-mismatch in acoustic scene classification.
\item We propose a knowledge distillation architecture for multi-target domain adaptation going from AST to MobileNets.
\end{enumerate}

\section{Related Works}

Domain-Adversarial Neural Networks~\cite{DANN} introduced the gradient reversal layer between the feature extractor and a
domain discriminator, encouraging a domain-invariant feature space. Gharib et al.~\cite{ASC_with_ADA} performs adversarial training between source and target in a stage-wise manner.
Drossos et al.~\cite{ASC_with_WGAN} follow ~\cite{ASC_with_ADA} and replace the discriminator objective with a Wasserstein distance between the domains. Olvera et al.~\cite{BlindDA_with_BN} used adversarial adaptation with input normalization and removed paired samples from the source and target recordings to remove the correlation between domains.

Prototypical networks~\cite{Prototypical} and feature disentanglement~\cite{disentanglement} treat the recording city as a possible domain shift. CliqueNets with
mixup~\cite{CliqueNets} instead make  the classifier robust against device mismatch through data augmentation. Device impulse response augmentation~\cite{DIR} simulates unseen devices from labeled source recordings and uses a similar pipeline and data split as this work.

Established UDA methods from vision still remain untested on audio spectrogram transformers and pretrained CNN models.
Maximum classifier discrepancy~\cite{MCD} align domains through the disagreement of two
classifier heads. Source Hypothesis transfer (SHOT) ~\cite{shot} drops the source data and adapts
under pseudo-labels. Transferable vision transformers (TVT)~\cite{TVT} exploits patch tokens and applies
only to transformers. To the best of our knowledge, the current work is the first
comparison of discrepancy-based, source-free, and
transformer-specific adaptation for device-invariant ASC.
%and the first
%to measure how much device information each method removes from a PaSST
%backbone.

Several works extend UDA to multiple target domains through explicit feature alignment. \cite{pei2018multi} introduces class-wise domain discriminators, which align source and target distributions at a fine-grained level. \cite{gholami2020unsupervised} disentangles a shared latent space from domain-specific factors. This shared space supports adaptation across multiple targets jointly. \cite{isobe2021multi} trains per-target expert models under a consistency constraint. These experts align predictions across target domains through a shared bridge.

Knowledge distillation offers an alternative to feature alignment. \cite{nguyen2021unsupervised} trains one teacher per target domain, distilling soft-label outputs into a shared student network. Distillation alternates across teachers over successive training epochs, transferring functional knowledge without any feature-level supervision.

For ASC, \cite{takeyama2021robust} adapted device-specific models using classifier discrepancy, and distill these specialists into one device-general ASC model.
Yang et al. ~\cite{MTDA} extend the setting to multiple target domains with a multi-class domain discriminator, and order the domains from easy to hard to adapt, constructing a weighted loss function for multiple-target domain adaptation.

\section{Methods for UDA}
\label{sec:method}

%We briefly describe below various UDA methods applied to ASC.

\subsection{Adversarial alignment}
MCD and DANN share similar objective functions. The feature extractor $G_f$
and  scene classifier $G_c$ are trained with a cross-entropy loss $\mathcal{L}_\text{cls}$ on labelled
source data, while an adversary $D_g$ tries to guess the source and target features apart. The objective function is as follows:
\begin{align}
    \min_{G_f, G_c} \;\;
    \mathcal{L}_{\text{cls}}\big(G_c(G_f(x_s)),\, y_s\big)
    \;+\; \lambda \max_{D_g} \;
    \mathcal{L}_{\text{adv}}.
\end{align}
MCD and DANN differ only in adversarial loss $\mathcal{L}_\text{adv}$ . DANN~\cite{DANN} uses a domain
discriminator trained through a gradient reversal layer, which aims to
strip domain information from the embedding. MCD~\cite{MCD} replaces the
discriminator with a second scene classifier and maximises the mean
absolute gap between the two softmax outputs.

\subsection{Source-free adaptation}
SHOT~\cite{shot} adapts to the target domain without access to source data.  The source trained classifier is frozen while the feature extractor $G_f$ is updated. The method optimizes two objectives. The first objective is an information-maximization term $\mathcal{L}_\textit{im}$, which sums an entropy loss and a diversity loss. The entropy loss tries to increase the confidence of each prediction, while the diversity loss encourages the outputs to differ from each other. The second objective is a cross-entropy term $\mathcal{L}_\textit{pl}$  computed between the predicted labels and pseudo labels obtained via k-means clustering. The final objective is
\begin{align}
   \mathcal{L}_\textit{obj} =   \mathcal{L}_\textit{im} + \mu\,\mathcal{L}_{\text{pl}}
\end{align}
 where $\mu$ is a hyperparameter.

\subsection{Transformer-specific adaptation}
TVT~\cite{TVT} exploits the patch tokens of a vision transformer, and we
adapt TVT to PaSST (described later.) Both the patch-level discriminator $D_l$ and the global discriminator $D_g$ are employed here. TVT uses a Transferability Adaptation Module (TAM). TAM scores every patch on its transferability using entropy from logits of $D_l$. These scores are further used to weight the attention of each patch. TVT also adds an information maximization term in the loss function.
We observed in our experiments that removing it costs nothing on the accuracy; hence, we haven't included it. The $\mathcal{L}_\textit{local}$ is the patch-level loss, whereas $\mathcal{L}_\textit{global}$ is the global loss applied on the CLS token. The updated loss function can be summarized as:
\begin{align}
    \min_{G_f, G_c} \;\;
    &\mathcal{L}_{\text{cls}}\big(G_c(G_f(x_s)),\, y_s\big)
    \; + \; \lambda \max_{D_g} \; \mathcal{L}_{\text{global}} \nonumber \\
    &+ \; \beta \max_{D_l} \; \mathcal{L}_{\text{local}}
\end{align}

\section{Multi-target Domain adaptation through Knowledge Distillation}
% For transferring knowledge to student models from teacher models, various algorithms have been proposed, which exclusively rely on ground-truth labels.
The teacher-student distillation framework uses teacher's soft probability targets to optimize the student. Let $D_S$ be a source domain, $D_{T_1}$, $D_{T_2}$ be target domains. We aim to develop a single student model which is invariant to all three domains using KD. For this, we
deploy two teacher models $T_1$ and $T_2$, respectively aligning $D_S$ to $D_{T_1}$ and $D_S$ to $D_{T_2}$. This is achieved using DANN or MCD.
%To assess the coalition's objectives, teachers are primarily pre-trained using DANN~\cite{DANN} and extended for MCD~\cite{MCD} method.
During KD, teachers remain frozen post-training. The models ${T_1}$ and $T_2$ use PaSST, and the student model employs MobilenetV3 architecture \cite{mobilenetv3}.

\textbf{Dynamic Domain Routing}:
Let $w_s \in D_S$ , $w_{T_1} \in D_{T_1}$, $w_{T_2} \in D_{T_2}$ be minibatches of audio spectrograms sampled concurrently and dynamically routed to generate domain-aligned soft targets:
\begin{align}
    q_s &= \frac{1}{2}\left\{\frac{T_1[w_s]}{K}+\frac{T_2[w_s]}{K}\right\},\\
    q_{T_1} &= \frac{T_1[w_{T_1}]}{K},\\
    q_{T_2} &= \frac{T_2[w_{T_2}]}{K}
    \label{eq:q_i}
\end{align}

Here, $q_s$ is the average logits of $T_1$ and $T_2$ when the input in the mini-batch is from $D_S$. Similarly, for input from $D_{T_1}$ and $D_{T_2}$ we take logits $q_{T_1}$ only from $T1$ and $q_{T_2}$ only from $T_2$, respectively. These logits are softened by temperature $K$ ($K>1$, softer targets, more inter-class information transferred; $K=1$ retains the original behavior).
The input mini-batches and routed soft targets are concatenated along the batch dimension as  $q_i = Concat(q_s,q_{T_1},q_{T_2})$ and $w_i = Concat(w_s,w_{T_1},w_{T_2})$  and given to the student with the following distillation loss:
\begin{align}
    \mathcal{L}_{\text{distill}} &= -\frac{K^2}{N} \times\sum_{i=1}^{N}\sum_{c=1}^{C} q_i \log(p_i).
    \label{eq:L_distill}
\end{align}
The student $f_s$ predicts probabilities $p_i = f_s(w_i)$ and cross entropy is minimized against $q_i$ over all $C$ classes and  batch size of $N$,  as in Equation \eqref{eq:L_distill}.

\section{Experiments and Results}

%In this section of the paper, we will discuss the dataset, and the implementation details as described above.%

The performance of the UDA methods and the proposed distillation method is demonstrated on Task 1a of DCASE 2020~\cite{2020dcase} and Task 1 of DCASE 2022~\cite{2022dcase} datasets (containing 10 sec and 1 sec examples respectively). There are three real devices (A, B, and C) and six simulated devices (S1-S6) and 10 classes ($C$) of  acoustic scenes. We use device A as the source domain  and devices B-S3 as the target domain. For MTDA using KD, device A is the source domain $D_S$ and $D_{T_1}$ and $D_{T_2}$ are respectively from devices B, C and S1, S2. For all experiments, accuracy of classification is used  as evaluation metric.

%\subsection{Feature Extractors}
We use the Patchout Spectrogram Transformer (PaSST)~\cite{Passt} as our feature extractor. PaSST discards patches at adaptation time making the training faster than a full AST without losing accuracy. Additionally, we consider a MobileNet-based model~\cite{mobilenetv3}, which is  pre-trained on AudioSet. %and further refined via knowledge distillation from PaSST.
In the MTDA setup, intermediate feature
representations from PaSST serve as soft supervision signals for the MobileNetV3. MobileNetV3 is much more efficient at inference time and serves as a compact CNN alternative to PaSST.

%for UDA studies as well as teacher models $T_1$ and $T_2$ for KD.

% MobileNet student, yielding a compact model that retains much of PaSST's
% representational capacity while being substantially more efficient at inference
% time.

\subsection{Implementation details}

The PaSST takes a 768-dimensional embedding as its input, which is passed to a fully connected layer of 128 units. These are then fed to an output
layer of 10 acoustic scene classes. The MobileNetV3 uses two fully connected layers as classifier. Its input is the 960-dimensional output of the MobileNetV3 feature extractor, mapped to a hidden layer of 1280 units. The hidden layer applies a Hardswish activation followed by 20\% dropout. The output layer then maps to the same 10 classes.

For UDA, gradients reach PaSST through its classification token. Every parameter
in the backbone updates during training. We optimize with Adam at a learning rate of $1\times10^{-5}$. The
classifier and the domain discriminator train at $1\times10^{-3}$. We
implement every method in PyTorch. Each implementation starts from the
code released by its original authors and are adapted for PaSST.
%TVT required changes to the PaSST architecture,   including addition of a Transferability Adaptation Module layer in the penultimate block, as mentioned in Section XXXX.
For MTDA, the GRL- and MCD-based adapted PaSST are used as teacher models to distill the MobilenetV3 based student model. The same settings are carried for the knowledge distillation task over 30 epochs for the softening parameter $K=2$.  All training runs on a single RTX Pro 5000 GPU.

\begin{table*}[htbp]
\caption{Device-to-device ASC accuracy (\%) on TAU Urban Acoustic Scenes 2020 (10\,s clips) and 2022 (1\,s clips). A$\rightarrow$A is the source-domain reference and needs no adaptation. Avg is the mean over the five unseen devices. TVT applies to transformers only.}
\centering
\resizebox{\textwidth}{!}{%
\begin{tabular}{l|c|cc|ccc|c||c|cc|ccc|c}
\toprule
 & \multicolumn{7}{c||}{\textbf{TAU 2020 (10\,s)}} & \multicolumn{7}{c}{\textbf{TAU 2022 (1\,s)}} \\
\textbf{Method} & \textbf{A$\to$A} & \textbf{A$\to$B} & \textbf{A$\to$C} & \textbf{A$\to$S1} & \textbf{A$\to$S2} & \textbf{A$\to$S3} & \textbf{Avg} & \textbf{A$\to$A} & \textbf{A$\to$B} & \textbf{A$\to$C} & \textbf{A$\to$S1} & \textbf{A$\to$S2} & \textbf{A$\to$S3} & \textbf{Avg} \\
\midrule
PaSST source & 82.5 & 66.0 & 70.2 & 60.0 & 56.6 & 61.0 & 62.76 & 70.45 & 47.9 & 60.27 & 27.7 & 23.03 & 30.6 & 37.9 \\
\; + DIR      & --   & 72.0 & 69.3 & 69.1 & 62.7 & 67.6 & 68.1  & --    & 52.1 & 60.7  & 42.2 & 41.9  & 44.8 & 48.3 \\
\; + DANN     & --   & 73.0 & 79.0 & \textbf{74.8} & \textbf{73.9} & \textbf{74.0} & \textbf{74.9} & -- & 54.9 & 64.4 & 47.0 & 45.8 & 53.3 & 53.1 \\
\; + MCD      & --   & 71.1 & 78.0 & 73.9 & 68.8 & 74.2 & 73.2  & --    & 52.7 & 63.4  & 47.5 & 42.8  & 48.3 & 50.9 \\
\; + SHOT     & --   & 72.3 & 79.6 & 60.9 & 57.0 & 68.8 & 67.7  & --    & 48.4 & 57.0  & 32.4 & 30.8  & 42.0 & 42.1 \\
\; + TVT      & --   & \textbf{74.8} & \textbf{82.1} & 70.0 & 66.1 & 72.1 & 73.0 & -- & 51.9 & 61.3 & 44.9 & 37.8 & 47.7 & 48.5  \\
\midrule
MobileNetV3 source & 83.0 & 53.5 & 72.0 & 40.9 & 37.6 & 42.4 & 49.28 & 65.3 & 40.7 & 56.6 & 25.7 & 25.3 & 26.6 & 35.0 \\
\; + DIR      & --   & 59.5 & 71.12 & 58.78 & 55.75 & 60.3 & 61.1  & --    & 40.5 & 53.3  & 37.1 & 36.1  & 39.21 & 41.25 \\
\; + DANN     & --   & \textbf{65.1} & 75.7 & 54.6 & \textbf{58.8} & \textbf{62.7} & \textbf{63.4} & -- & \textbf{48.2} & \textbf{59.6} & \textbf{40.8} & 39.0 & 45.6 & \textbf{46.6} \\
\; + MCD      & --   & 62.9 & \textbf{77.2} & \textbf{56.1} & 55.1 & 61.2 & 62.5 & -- & 48.8 & 58.8 & 38.4 & \textbf{40.4} & 42.4 & 45.7 \\
\; + SHOT     & --   & 59.7 & 72.9 & 52.7 & 50.0 & 60.9 & 59.3  & --    & 46.3 & 56.2  & 37.8 & 37.5  & \textbf{47.0} & 45.0 \\
\bottomrule
\end{tabular}%
}
\label{tab:adaptation_results}
\end{table*}

\begin{figure}[t]
%\centerline{\rule{\linewidth}{4cm}} % This generates a placeholder black box
\centering
\includegraphics[width=\linewidth]{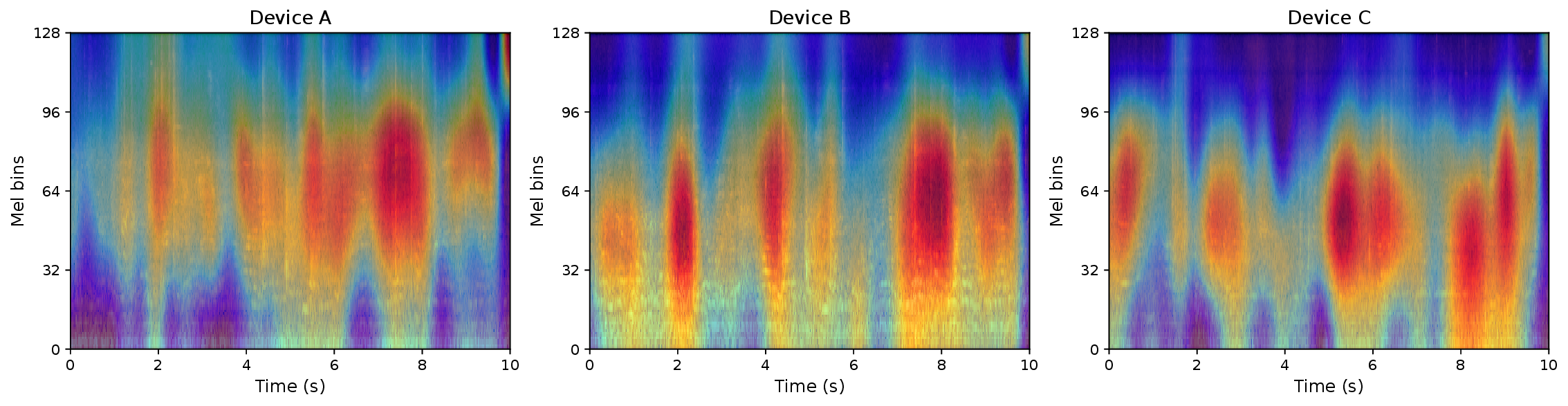}
\caption{Grad-CAM visualization of activations on the distilled student model Student-GRL for the same recording across three devices A, B and C. The similar activations indicate the student model is invariant across devices.}

\label{fig:ABC_2020}
\end{figure}

% \begin{figure}[t]
% %\centerline{\rule{\linewidth}{4cm}} % This generates a placeholder black box
% \centering
% \includegraphics[width=\linewidth]{ABC_TAU2020png.drawio.png}
% \caption{Performance comparison between PaSST, MobilenetV3 Baselines and distilled Student $S1$ using GRL based teacher on Device A, B, C  for TAU 2020 dataset.}
% \label{fig:ABC_2020}
% \end{figure}

%\vspace{-0.6cm}
\subsection{Baseline: source model}

We finetune PaSST on the training portion of data from Device A. Testing the
finetuned model on the test portion of Device A gives the \textbf{matched-setting} or \textbf{source baseline} of 82.5\% on TAU 2020. Testing the same model on B, C, S1, S2, and S3 gives the \textbf{mismatched setting}, where accuracy drops 19\% on average. Device-wise figures are mentioned in Table~\ref{tab:adaptation_results}. The drop clearly indicates a domain shift between devices.
We repeat the experiments with MobileNetV3. The matched-setting baseline is 83\% on TAU 2020. On target devices we observe an average drop of 34\%. We observe that the domain shift is larger in the case of MobileNetV3.
Table~\ref{tab:adaptation_results} also reports results on TAU 2022, which are shorter recordings.
PaSST's source baseline of 70.45 drops by 32\% in the  mismatched setting. Similarly, MobileNetV3 sees an accuracy drops of  30.3\% in the mismatched setting.

\subsection{Domain adaptation}

Table~\ref{tab:adaptation_results}  reports the results of various UDA methods. Every method improves on the mismatched baseline for both backbones.
% On the TAU 2020 dataset, the mismatched-condition achieves an average accuracy of 62.7\% across the five unseen devices when using PaSST as the feature extractor, compared with 49.2\% for MobileNetV3.
On TAU 2020, DANN yields the largest improvement for both feature extractors, increasing the average accuracy for all devices by 12.2\% for PaSST and 14\% for MobilenetV3. MCD gives improvements of 11\% and 13\%, whereas SHOT gives improvement of 5\% and 10\%. TVT gives an improvement of 10.3\% on PaSST. Similar trends are seen for TAU 2022 as well, with similar relative gains.

%to 75.0\% with PaSST and 63.4\% with MobileNetV3. MCD achieves 73.2\% and 62.5\% with PaSST and MobileNetV3, respectively, while TVT attains 72.7\% with PaSST. SHOT achieves 67.7\% with PaSST and 59.3\% with MobileNetV3.
%On TAU 2022, MobileNetV3 achieves an average accuracy of 46.6\% with DANN, 45.7\% with MCD, and 45.0\% with SHOT. For PaSST, the average accuracy is 53.1\% with DANN, 50.9\% with MCD, 48.5\% with TVT, and 42.1\% with SHOT.

For comparison, we include the Device Impulse Response Augmentation (DIR) \cite{DIR}  results under the mismatched device setting. DIR achieves an average gain of 9\% for both datasets across both pipelines, and remains an alternative method for overcoming domain shift.

%acccuracy of 68.1\%  using PaSST as the feature extractor on Tau 2020 and 48.3\% on Tau 2022. For MobileNetV3, DIR achieves an average of 61.1\% on TAU 2020 and 41.25\% on TAU 2022.

\subsection{Multi-Target Domain Adaptation}
\begin{figure}[t]
\centering
\includegraphics[width=\linewidth]{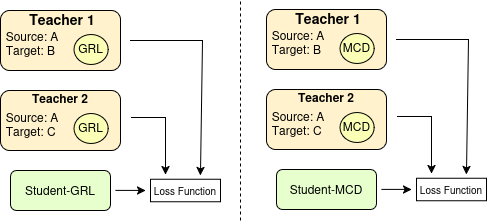}
\caption{KD architecture. Student-GRL is distilled from GRL based teachers and Student-MCD is distilled from MCD based teachers. Source domain is device A, and targets can be devices B and C, or devices S1 and S2. See text for details.}
\label{fig:KD_method}
\end{figure}

Table ~\ref{tab:tau_stacked} reports the accuracy of the student MobileNetV3 model distilled from two teacher models, each aligned from the source domain to different target domains. Here the source domain $D_S$ is device A, and the target domains $D_{T_1}$ and $D_{T_2}$ are devices B and C respectively. Domain adaptation is performed using either GRL or MCD and then distilled into the student model, resulting in models  Student-GRL and Student-MCD (see Figure \ref{fig:KD_method}).

In Table \ref{tab:tau_stacked}, the ``Source'' row gives the accuracies for matched and mismatched condition on  the original MobileNetV3 without distillation. The remaining rows give the accuracy of the distilled student MobileNetV3, exhibiting improved generalization across devices A, B and C. There is an average improvement of 7\% for TAU 2020 and 4\% for TAU 2022.

We additionally distilled another student MobileNetV3 model from two teachers adapted from domains A to S1 and A to S2. The resulting student generalises well on domains A, S1 and S2, with an average improvement of 25\% on TAU 2020 and 11\% on TAU 2022, over the original MobileNetV3.

%Student-GRL achieves an average classification accuracy of 72.76\% on devices B and C for TAU 2020, while Student-MCD achieves 68.24\%. For TAU 2022, Student-GRL and Student-MCD achieve average accuracies of 52.85\% and 53.23\%, respectively.

These results suggest that domain-specific knowledge
is successfully distilled from multiple teachers into a single student. Both GRL
and MCD therefore perform successful domain adaptation, which is
used in the KD phase.

%distilled MobileNetV3 studnets on source domain A and the two  students on source domain A and the two teacher target domains B and C.
%On TAU 2020, Student-GRL reaches 68.39\% on B and 77.20\% on C, a gain of 14.9 and 5.2 points over the MobileNetV3 source-only baseline for Device B and C.  Student-MCD gains 10.6 points on B (64.13\% vs. 53.5\%) and remain approximately same on C.  MCD lags as a teacher when compared to DANN.  Both students hold source-domain accuracy at the 83.0\% baseline (82.12\%, 83.03\%), so distillation costs nothing on A.

%On TAU 2022, Student-GRL shows the same pattern: 50.57\% on B and 55.13\% on C sit 4.3 and 9.3 points below the corresponding PaSST-DANN teachers, while A improves on the 65.3\% baseline to 67.39\%. XXX Need to add MCD Tau 2022 XXX

% \begin{table}[htbp]
%   \centering
%   \caption{Knowledge Distillation performance on real devices (A, B, C) across TAU 2020 and TAU 2022. Students $S1$ and $S2$ are distilled from GRL and MCD based teachers, respectively.}
%   \label{tab:knowledge_distillation_real}
%   \begin{tabular}{l|cccccc|cccccc}
%     \toprule
%     & \multicolumn{3}{c|}{\textbf{TAU 2020}} & \multicolumn{3}{c}{\textbf{TAU 2022}} \\
%     \textbf{Student} & \textbf{A} & \textbf{B} & \textbf{C} & \textbf{S1} & \textbf{S2} & \textbf{S3} & \textbf{A} & \textbf{B} & \textbf{C} & \textbf{S1} & \textbf{S2} & \textbf{S3}\\
%     \midrule
%     S1 (GRL) & 82.12 & 68.38 & 77.20 & 67.39 & 50.57 & 55.13 \\
%     S2 (MCD) & 83.03 & 64.13 & 72.34 & a & b & c \\
%     \bottomrule
%   \end{tabular}
% \end{table}

\begin{table}[t]
\centering
\caption{Distilled MobileNetV3 student accuracy (\%) on devices A, B, C. Source row indicates performance of original MobileNetV3 without any distillation. Student-GRL and Student-MCD are distilled from PaSST teachers adapted with DANN and MCD, respectively. The student models show improved cross-domain accuracy.}
\label{tab:tau_stacked}
\begin{tabular}{ll ccc}
\toprule
\textbf{Dataset} & \textbf{MobilenetV3} & \textbf{A} & \textbf{B} & \textbf{C} \\
\midrule
\multirow{3}{*}{TAU 2020}
& Source      & 83.0  & 53.5  & 72.0  \\
& Student-GRL & 82.12 & 68.39 & 77.20 \\
& Student-MCD & 83.03 & 64.13 & 72.34 \\
\midrule
\multirow{3}{*}{TAU 2022}
& Source      & 65.3  & 40.7  & 56.6  \\
& Student-GRL & 67.39 & 50.57 & 55.13 \\
& Student-MCD & 68.39 & 46.39 & 60.06 \\
\bottomrule
\end{tabular}
\end{table}

\begin{figure}[t]
    \centering
    \includegraphics[width=0.9\columnwidth]{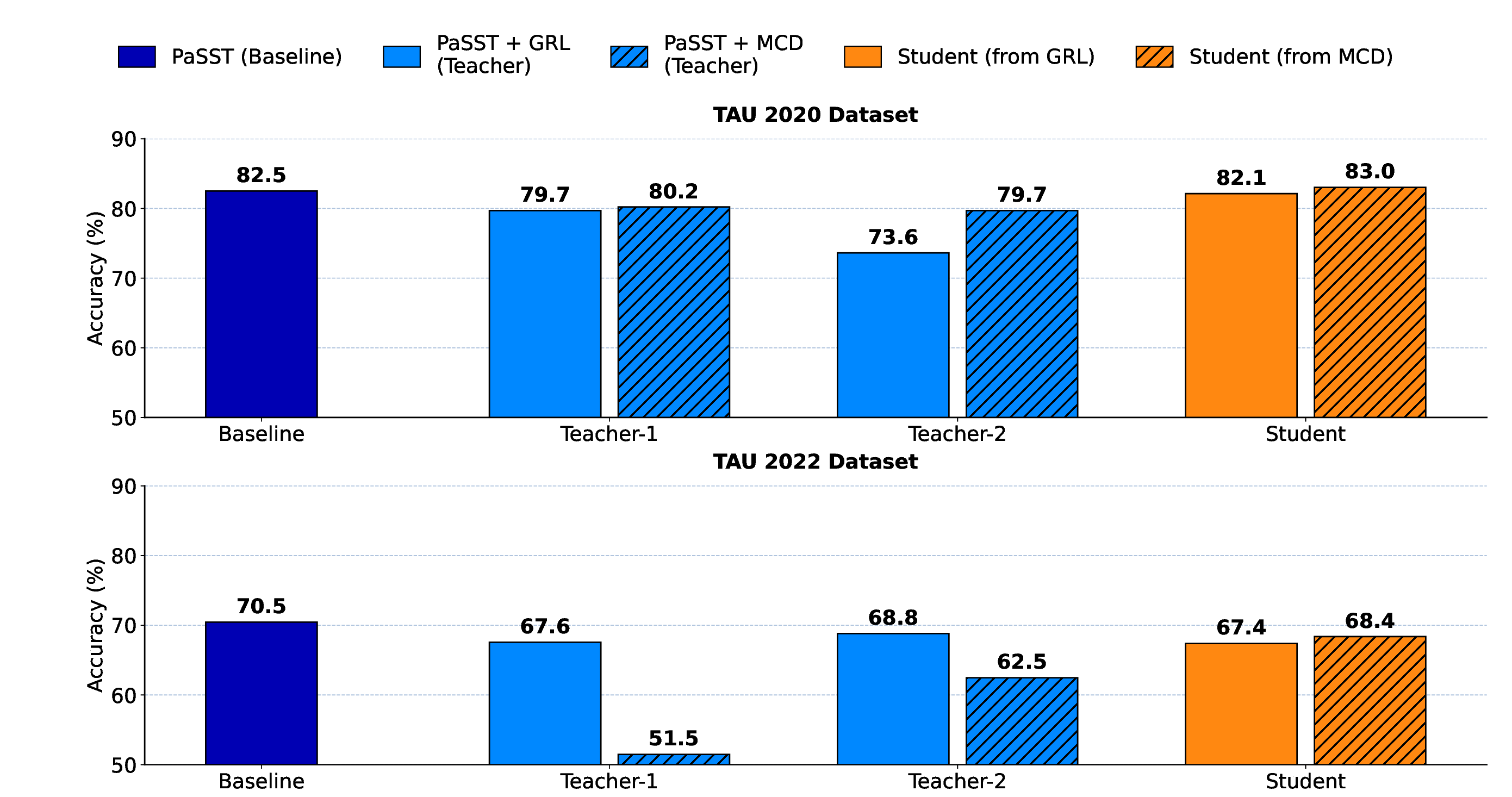}
    \caption{The baseline bar shows accuracy of PaSST model trained and tested on domain A. Teacher models $T_1$ and $T_2$ show a drop in performance on domain A. Whereas the distilled student retains performance on domain A.
    %    Comparison between Device A classification accuracy of single-target adapted teachers ($T_1$ and $T_2$) using both domain adaptation methods and distilled students on TAU 2020 (top) and TAU 2022 (bottom) datasets.
    }
    \label{fig:A_comp}
\end{figure}

\section{Discussion}
\subsection{Methods for domain adaptation}
It is worth noting that, although DANN \cite{DANN} is an older domain adaptation method and has generally been outperformed by more recent approaches including MCD, TVT, and SHOT in the vision domain, this observation does not appear to extend to the audio domain. In our experiments, DANN performs comparatively well when applied to audio spectrograms, despite its relatively simple adaptation mechanism.

We additionally experiment on other transformer-specific methods such as CDTRANS \cite{cdtrans} and cycle self-training \cite{CST}. Unlike in vision, these methods were unable to overcome the domain shift from source to target domain for spectrograms. At this point of time, we are unable to explain this observation.

%Figure \ref{fig:ABC_2020} shows GRAD-CAM \cite{gradCam} activations for MobileNet for two domains before and after GRL-based adaptation. It can be seen that activations across devices are consistent after adaptation, showing
%domain-invariant features.

% ----------------- should i add this ~~~~~~~~~~%
We observe that PaSST benefits more than MobileNetV3 from domain adaptation with the most gains in the simulated devices across methods.
The shorter 1-second clips in TAU 2022 result in a substantially more challenging mismatched setting.
%For PaSST, the source-only average accuracy decreases from 62.76\% on TAU 2020 to 37.9\% on TAU 2022.
Although DANN improves the average accuracy of TAU 2022 to 53.1\%, its performance remains below that obtained on the TAU 2020. This indicates that domain adaptation reduces the device mismatch but does not fully compensate for the reduced temporal context.

% Figure~\ref{fig:ABC_2020} shows Grad-CAM~\cite{gradCam} activations for MobileNetV3 before and after GRL-based adaptation on devices A and B. Before adaptation, the activation patterns differ substantially across the two devices, whereas after adaptation, the model exhibits more similar activation patterns for recordings from the two devices. This suggests that GRL-based adaptation changes the model's attention toward more device-invariant regions.

Figure~\ref{fig:ABC_2020} shows the Grad-CAM~\cite{gradCam} visualization of the distilled student model obtained using Teacher models $T_1$ and $T_2$. The sample corresponds to the same acoustic scene recorded simultaneously using devices A, B and C respectively. The highlighted regions remain concentrated around similar time–frequency structures across all devices. This consistency suggests that the distilled student model focuses on similar acoustic patterns and discriminative regions despite device-induced variations, indicating that the learned representations are relatively robust to domain shifts introduced by differences in recording devices.

% The results also show that data augmentation method (DIR) alone provides a  improvement over the matched condition. DIR increases the average PaSST accuracy to 68.1\% on TAU 2020 and 48.3\% on TAU 2022. The learned domain adaptation methods achieve higher average accuracies. This indicates that explicitly learning to reduce the source target discrepancy can provide benefits beyond those obtained by simulating device characteristics through augmentation.

\subsection{Retaining of domain information after knowledge distillation}

We also explore if the PaSST-based teacher models forget the source domain after adaptation. Figure \ref{fig:A_comp} shows the source domain accuracy of the two teacher models $T_1$ and $T_2$ after adaptation. It is seen that there is a drop in accuracy in the source domain. On the other hand, the student model distilled from the two teachers retains more of the source domain information (domain A), in addition to being invariant to unseen domains B and C. This implies that knowledge distillation does not forget the original domain, especially in the case of TAU 2020.

\section{Conclusion and Future Work}

This paper compares existing unsupervised domain adaptation methods for acoustic scene classification across two backbones, two datasets, and five device shifts. The PaSST backbone provides more domain resilience than MobileNetV3. Domain adaptive neural network works well across most devices, whereas transferable vision transformer, maximum classifier discrepancy and source free hypothesis transfer give encouraging performance. In general, the short recordings of TAU 2022 are more challenging for domain adaptation.

Additionally, we propose a knowledge distillation based approach for multi-target domain adaptation. We use multiple domain-adapted teachers to distill information into a student model. The student model demonstrates better domain retention, and improved domain generalization.

\bibliographystyle{IEEEtran}
\bibliography{refs}

\end{document}